\documentclass[pdflatex,sn-nature]{sn-jnl}

\usepackage{graphicx}%
\usepackage{multirow}%
\usepackage{amsmath,amssymb,amsfonts}%
\usepackage{amsthm}%
\usepackage{mathrsfs}%
\usepackage[title]{appendix}%
\usepackage{bm}
\usepackage{braket}
\usepackage{gensymb}
\usepackage{xcolor}%
\usepackage{textcomp}%
\usepackage{manyfoot}%
\usepackage{booktabs}%
\usepackage{algorithm}%
\usepackage{algorithmicx}%
\usepackage{algpseudocode}%
\usepackage{listings}%
\UseRawInputEncoding

\DeclareMathOperator*{\bdg}{BdG}

\DeclareMathOperator*{\tr}{Tr}

\theoremstyle{thmstyleone}%
\theoremstyle{thmstyletwo}%

\theoremstyle{thmstylethree}%

\begin{document}

\title[Article Title]{Field-Tunable Anisotropic Fulde-Ferrell Phase in NbSe$_2$/CrSiTe$_3$ Heterostructures}


\author[1,2]{\fnm{Jiadian} \sur{He}}
\equalcont{These authors contributed equally to this work.}

\author[2]{\fnm{Xin-Zhi} \sur{Li}}
\equalcont{These authors contributed equally to this work.}
\author[1,2]{\fnm{Chen} \sur{Xu}}
\author[1,2]{\fnm{Yifan} \sur{Ding}}

\author[1,2]{\fnm{Yueshen} \sur{Wu}}
\author[1,2]{\fnm{Jinghui} \sur{Wang}}
\author[1,2]{\fnm{Peng} \sur{Dong}}

\author[2]{\fnm{Yan-Fang} \sur{Li}}

\author[3]{\fnm{Wei} \sur{Li}} \email{w\_li@fudan.edu.cn}

\author*[1,2]{\fnm{Xiang} \sur{Zhou}}\email{zhouxiang@shanghaitech.edu.cn}
\author[1,2]{\fnm{Yanfeng} \sur{Guo}}
\author[1,2,4]{\fnm{Yulin} \sur{Chen}}

\author*[2]{\fnm{Wen-Yu} \sur{He}}\email{hewy@shanghaitech.edu.cn}

\author*[1,2]{\fnm{Jun} \sur{Li}}\email{lijun3@shanghaitech.edu.cn}

\affil[1]{\orgdiv{ShanghaiTech Laboratory for Topological Physics}, \orgname{ShanghaiTech University}, \orgaddress{\city{Shanghai}, \postcode{201210}, \country{China}}}

\affil[2]{\orgdiv{State Key Laboratory of Quantum Functional Materials, School of Physical Science and Technology}, \orgname{ShanghaiTech University}, \orgaddress{\city{Shanghai}, \postcode{201210}, \country{China}}}

\affil[3]{\orgdiv{State Key Laboratory of Surface Physics and Department of Physics}, \orgname{Fudan University}, \orgaddress{\city{Shanghai}, \postcode{200433}, \country{China}}}

\affil[4]{\orgdiv{Department of Physics, Clarendon Laboratory}, \orgname{University of Oxford}, \orgaddress{\city{Oxford}, \postcode{OX1 3PU}, \country{United Kingdom}}}


\abstract{The emergence of superconductivity in two-dimensional transition metal dichalcogenides with strong spin orbit coupling (SOC) has opened new avenues for exploring exotic superconducting states. Here, we report experimental observation of an anisotropic Fulde-Ferrell (FF) phase in few-layer NbSe$_2$/CrSiTe$_3$ heterostructures under in-plane magnetic fields. Through combined magnetoresistance and nonreciprocal transport measurements, we find that due to the couplings from the ferromagnetic CrSiTe$_3$, a half-dome-shaped region emerges in the magnetic field-temperature ($B$-$T$) diagram. Importantly, the half-dome-shaped region exhibits finite second harmonic resistance with in-plane anisotropy, indicating that the superconducting state is an anisotropic FF phase. Through a symmetry analysis combined with mean field calculations, we attribute the emergent anisotropic FF phase to the CrSiTe$_3$ layer induced Rashba SOC and three-fold rotational symmetry breaking. These results demonstrate that heterostructure stacking is a powerful tool for symmetry engineering in superconductors, which can advance the design of quantum devices in atomically thin superconducting materials.}




\maketitle

\section{Introduction}\label{sec1}

In a conventional Bardeen-Cooper-Schrieffer (BCS) superconductor, electrons with opposite momenta condense into Cooper pairs, leading to a macroscopic phase-coherent condensation of Cooper pairs with zero net momentum~\cite{Tinkham01,Annett}. While the classic BCS theory of superconductivity has achieved remarkable success in superconductors preserving both the time reversal and inversion symmetries, breaking these symmetries unlocks exotic superconducting regimes. For instance, breaking the time reversal symmetry by an external magnetic field via the Zeeman effect can stabilize the Fulde-Ferrel-Larkin-Ovchinnikov (FFLO) state~\cite{Fulde01,Larkin01}, where Cooper pairs acquire a finite center-of-mass momentum. In noncentrosymmetric superconductors, the intrinsic spin orbit coupling (SOC) arising from the inversion symmetry breaking alters the symmetry of Cooper pairs~\cite{Sigrist02}, yielding a wealth of unconventional superconducting phenomena such as Ising superconductivity with an enhanced paramagnetic limiting field~\cite{Sigrist01, Jianting, Fai01, Saito01, Barrera}, mixed singlet-triplet pairing~\cite{Gorkov, Mazin}, and spontaneous nematic order~\cite{Pribiag, Rolf01, Yuhang01}. It is conceivable that the simultaneous break of time reversal and inversion symmetries can further enrich the manifold of unconventional pairing states that have a finite center-of-mass momentum~\cite{Gorkov02, Sigrist03, Kaur, Feigelman, NoahYuan}. Recent advances in dual electrostatic gating, van der Waals stacking, and heterostructure design in two-dimensional (2D) quantum materials have enabled the engineering of crystalline symmetry breaking in a more controllable way~\cite{Justin02, Fatemi, Sajadi, Geim, Cuizu, Liljeroth, Hueso, JunLi01, He2025, Ding2025, Jeong, Stevan02, Stevan01, Yiran01}. Such developments in the 2D material engineering provide a great opportunity to experimentally investigate how engineered symmetry breaking reshapes the finite-momentum pairing states.

Hexagonal NbSe$_2$ is a superconducting member of the transition metal dichalcogenide (TMD) family that has exhibited a range of intriguing superconducting properties~\cite{Jianting, Fai01, Saito01, Barrera, Pribiag, Rolf01, Yuhang01, Fai02, Jianwang, Jianhao, Jianming, Smet, PuhuaWan,Shuyun, Xiangfeng}. The bulk NbSe$_2$ has been widely considered as a standard $s$-wave superconductor~\cite{Fletcher}, whereas its superconducting properties undergo a profound evolution when the bulk NbSe$_2$ is thinned to the 2D limit~\cite{Fai01,Fai02,Jianwang,Pribiag,Rolf01}. For the few-layer NbSe$_2$ with atomic thickness, the broken in-plane inversion symmetry generates an Ising SOC that pins the electron spins to the out-of-plane direction, leading to Ising superconductivity with the hallmark that the in-plane upper critical field ($B_{\textrm{c}2}$) exceeds the Pauli paramagnetic limit ($B_{\textrm{P}}$)~\cite{Fai01, Fai02, Jianwang}. Recently, despite the three-fold rotational symmetry held by the lattice structure, the few-layer NbSe$_2$ has been observed to exhibit an emergent two-fold anisotropy in the in-plane $B_{\textrm{c}2}$, suggesting a spontaneous nematic pairing~\cite{Pribiag,Rolf01}. According to conventional FFLO theory, near the $B_{\textrm{P}}$, Cooper pairs tend to condense with a finite center-of-mass momentum~\cite{Fulde01,Larkin01}. Although for the few-layer NbSe$_2$, the in-plane $B_{\textrm{c}2}$ has been identified to exceed $B_{\textrm{P}}$ with an emergent two-fold rotational symmetry, experimental investigations on the possible finite momentum pairing under large in-plane magnetic fields remain limited. Crucially, the crystalline symmetry of few-layer NbSe$_2$ can be further manipulated through heterostructure design~\cite{Cuizu,Liljeroth,Hueso, JunLi01}, making it an unprecedented platform for unraveling the interplay between exotic pairing states, finite momentum pairing, and engineered symmetry breaking.

In this work, we present the experimental evidence of an anisotropic Fulde-Ferrell (FF) phase in few-layer NbSe$_2$/CrSiTe$_3$ heterostructures, by applying in-plane magnetic fields with varying directions. Initially, we observed that in NbSe$_2$/CrSiTe$_3$ heterostructures, the magnetoresistance (MR) near the superconducting phase transition exhibits non-monotonic behaviors, in contrast to the monotonic drop to zero as seen in pristine few-layer NbSe$_2$ samples. In the $B$-$T$ phase diagram of NbSe$_2$/CrSiTe$_3$ heterostructures, such anomalous MR curve occurs at low temperatures and high magnetic fields, which forms a half-dome-shaped region. Subsequently, we checked the nonreciprocal transport under in-plane magnetic fields and observed finite second harmonic resistance in the newly emergent half-dome-shaped region, indicating an FF phase superconducting state here. Finally, the in-plane anisotropy of the FF phase is revealed through two complementary observations: 1) temperature-dependent MR measurements showing anisotropic in-plane upper critical fields in perpendicular directions, and 2) anisotropic second-harmonic resistance under in-plane magnetic fields. In our experiment, the half-dome feature is absent in misaligned NbSe$_2$/CrSiTe$_3$ heterostructures, indicating that the alignment between the CrSiTe$_3$ and top NbSe$_2$ layer is essential for observing the FF phase. This alignment requirement underscores the crucial role of the coupling effects from the CrSiTe$_3$ layer. Inspired by these findings, we developed a theoretical model incorporating symmetry-reducing effects from CrSiTe$_3$. With a mean field calculation performed on the pairing states in NbSe$_2$ layers, we showed that the CrSiTe$_3$ induced three-fold rotational symmetry breaking plays a pivotal role in generating the anisotropic FF phase observed in the experiment.


\section{Results}\label{sec2}

\subsection{Device design and the superconductivity in NbSe$_2$/CrSiTe$_3$ heterostructures}\label{subsec1}

\begin{figure*}[htbp]
	\centering
	\includegraphics[width=1\textwidth,clip]{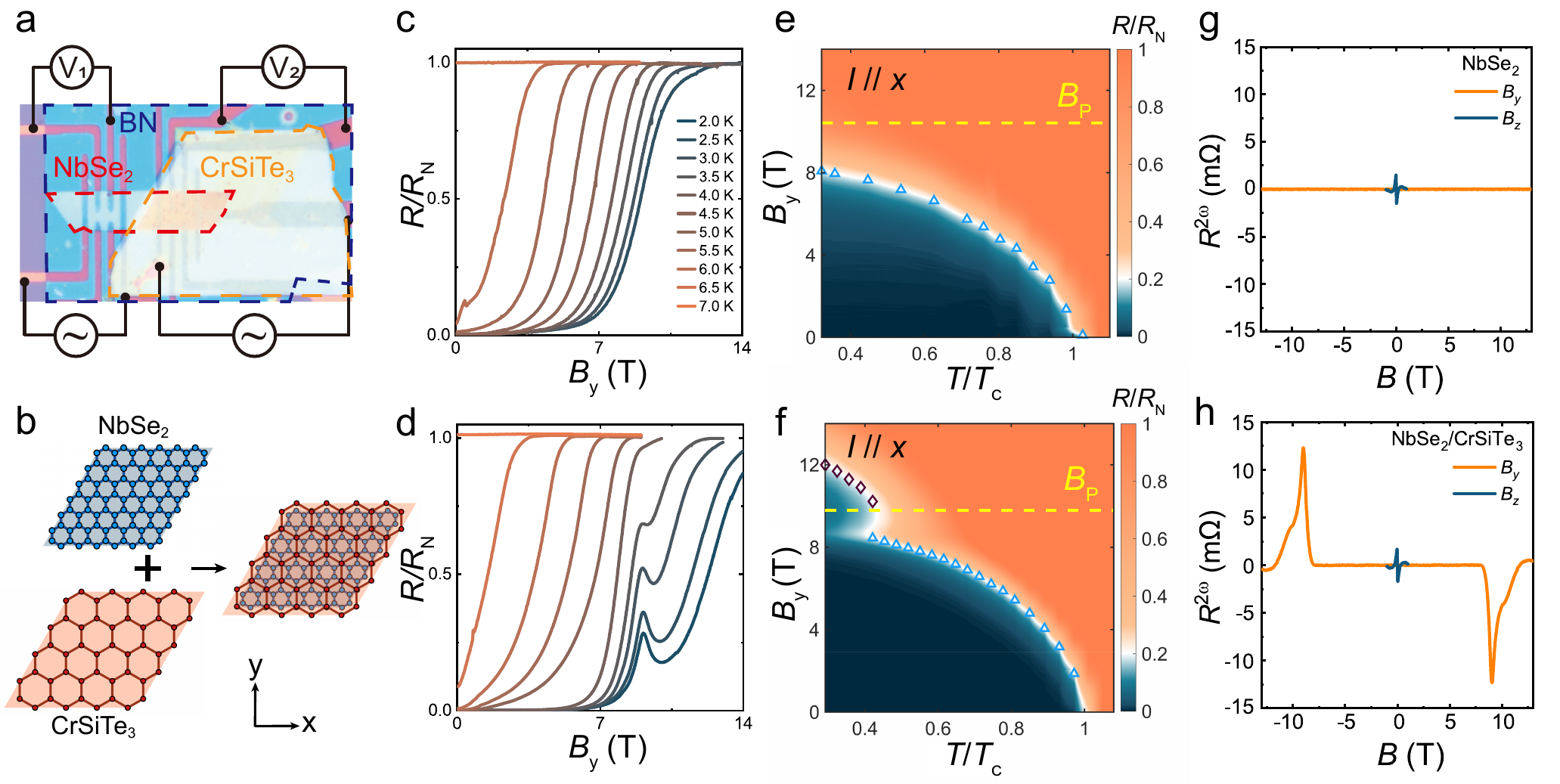}
	\caption{\textbf{Superconductivity and nonreciprocal transport properties in pristine NbSe$_2$ and NbSe$_2$/CrSiTe$_3$ heterostructures.} \textbf{a}, Optical image of the device structure for electrical transport measurement. The voltages signal of V$_1$ and V$_2$ correspond to the pristine NbSe$_2$ and NbSe$_2$/CrSiTe$_3$ heterostructures. The red, orange, and navy blue dashed boxes in the figure correspond to the NbSe$_2$, CrSiTe$_3$, and \textit{h}-BN thin flakes, respectively, with NbSe$_2$ having a thickness of 9 nm. \textbf{b}, The top-view crystal structures of monolayer NbSe$_2$ and CrSiTe$_3$, as well as their stacked crystal structure. \textbf{c}, \textbf{d}, The evolution of R-H
	curves at various temperatures for pristine NbSe$_2$ and NbSe$_2$/CrSiTe$_3$ heterostructures. \textbf{e}, \textbf{f}, The corresponding $B$-$T$ phase diagrams of the normalized magnetoresistance $R/R_\textrm{N}$ for pristine NbSe$_2$ and NbSe$_2$/CrSiTe$_3$ heterostructures, respectively. The yellow dashed lines represent the Pauli limit.  During the measurement, the magnetic field is applied along the \textit{y}-direction. Here the superconducting transition temperature at zero magnetic field is $T_{\textrm{c}}=6.1$ K. \textbf{g}, \textbf{h}, The second-harmonic magnetoresistance $R^{2\omega}$ of pristine NbSe$_2$ and NbSe$_2$/CrSiTe$_3$ heterostructures under in-plane and out-of-plane magnetic fields at $T$=3 K.}\label{fig1}
\end{figure*}

To investigate the influence of symmetry breaking in the NbSe$_2$/CrSiTe$_3$ heterostructure, we designed a control sample (Fig. \ref{fig1}\textbf{a}). The device was fabricated by first dry-transferring an elongated NbSe$_2$ strip (9 nm thick) onto pre-patterned electrodes, followed by the deterministic placement of a CrSiTe$_3$ flake ($\sim$ 40 nm) covering half of its length. The resulting heterostructure was then fully encapsulated with \textit{h}-BN. The schematic in Fig. \ref{fig1}\textbf{b} illustrates the final structure, with the long edges of the two flakes aligned.

Figures \ref{fig1}\textbf{c} and \ref{fig1}\textbf{d} show the MR curves measured at different temperatures in pristine NbSe$_2$ and NbSe$_2$/CrSiTe$_3$ heterostructures, respectively. In pristine NbSe$_2$, the MR decreases monotonically with reducing magnetic field, indicating a smooth transition into the superconducting state. In sharp contrast, while NbSe$_2$/CrSiTe$_3$ heterostructures exhibit a similar monotonic MR curve at $T>4$ K, below 3.5 K, an anomalous dip emerges in the MR curve. In the  magnetic field-temperature ($B$-$T$) diagrams shown in Fig. \ref{fig1}\textbf{e} and \ref{fig1}\textbf{f}, the non-monotonic MR curve of NbSe$_2$/CrSiTe$_3$ heterostructures yields an additional half-dome region of low-MR in the upper-left corner. Traced by brown diamonds and blue triangles for different regions, we find that $B_{\textrm{c}2}$ exceeds $B_{\textrm{P}}$ at low temperatures. Here, the $B_{\textrm{c}2}$ is determined at the 20\% of the normal-state resistance ($R/R_\textrm{N} = 0.2$). A comparison of Fig. \ref{fig1}\textbf{e} and \ref{fig1}\textbf{f} suggests that the newly emergent half-dome region in Fig. \ref{fig1}\textbf{d} originates from the interfacial coupling between NbSe$_2$ and CrSiTe$_3$ in the heterostructure.

At the interface of NbSe$_2$ and CrSiTe$_3$, the stacking of different crystalline structures predominantly breaks the $z$-directional inversion symmetry, which generates a strong Rashba SOC~\cite{Jiang2020}. For noncentrosymmetric materials, symmetry analysis has demonstrated that applying a magnetic field can induce nonreciprocal transport phenomenon~\cite{Rikken2001, Wakatsuki2017, LiYan2021, Yasuda2019, WuYueshen2022, Masuko2022, Wakatsuki2018, Itahashi2020, ZhangEnze2020}, where the corresponding nonreciprocal resistance $R$ is given by the phenomenological formula~\cite{Rikken2001, Wakatsuki2017, LiYan2021}:
\begin{equation}
    R = R_{0}[1+\gamma (\bm{B} \times \bm{P})\cdot \bm{I}]. \label{eq:first_harmo}
\end{equation}
Here $\bm{P}$ is the polarization vector denoting the inversion symmetry breaking, $\bm{I}$ is the applied current, and $\gamma$ denotes the nonreciprocal transport coefficient. Since the nonreciprocal resistance $R$ in Eq. \ref{eq:first_harmo} exhibits pronounced directional dependence with respect to both current and magnetic field orientations, we carried out nonreciprocal transport measurements to check the inversion symmetry breaking in the NbSe$_2$/CrSiTe$_3$ heterostructure.

According to Eq. \ref{eq:first_harmo}, the nonreciprocal resistance exhibits current-dependent nonlinearity: upon applying an AC current, the second harmonic resistance takes the form $R^{2\omega}=R_0\gamma\left(\bm{B}\times\bm{P}\right)\cdot\bm{I}^\omega$. Since $R^{2\omega}$ is proportional to $\bm{B}\times\bm{P}$, the direction of the polarization $\bm{P}$ that denotes the inversion symmetry breaking can be determined by measuring $R^{2\omega}$ via varying the direction of the applied magnetic field. Figures \ref{fig1}\textbf{g} and \ref{fig1}\textbf{h} show the in-plane ($B_y$) and out-of-plane ($B_z$) field dependent $R^{2\omega}$ measured at $T=3$K in pristine NbSe$_2$ and NbSe$_2$/CrSiTe$_3$ heterostructures, respectively. In the measurement, the AC current is applied along the $x$-direction. In both pristine NbSe$_2$ and the NbSe$_2$/CrSiTe$_3$ heterostructure, anti-symmetric $R^{2\omega}$ peaks emerge under small out-of-plane magnetic fields, which confirms the presence of weak intrinsic Ising SOC in NbSe$_2$~\cite{Wakatsuki2017, ZhangEnze2020}. In contrast, when the magnetic field is applied along the $y$-direction, a distinct anti-symmetric $R^{2\omega}-B_y$ signal emerges exclusively in the NbSe$_2$/CrSiTe$_3$ heterostructure, with no corresponding signal in pristine NbSe$_2$. This observation indicates that a strong Rashba SOC arises from the $z$-directional inversion symmetry breaking at the heterostructure interface~\cite{Yasuda2019, Itahashi2020, LiYan2021}. Furthermore, we calculated the nonreciprocal transport coefficient $\gamma$ by the equation $\gamma = \frac{2R^{2\omega}}{BI^{\omega}R^{\omega}} $ and find that the maximum $\gamma$ is much larger than other reported non-superconducting systems such as the Bi helix ($\sim$10$^{-3}$  T$^{-1}$ A$^{-1}$)~\cite{Rikken2001} and chiral organic materials ($\sim$10$^{-2}$  T$^{-1}$ A$^{-1}$)~\cite{pop2014}. For comparison, we also measured the $R^{2\omega}$ in a NbSe$_2$/Gr/CrSiTe$_3$ heterostructure and found that graphene intercalation suppresses the $R^{2\omega}-B_y$ signal (see Supplementary Note 2 and Supplementary Fig. 4). The suppressed nonreciprocal transport results further verify the presence of a Rashba type SOC at the NbSe$_2$/CrSiTe$_3$ interface. 



Having confirmed the Rashba type SOC at the NbSe$_2$/CrSiTe$_3$ interface, we now analyze the superconducting state within the emergent half-dome region in the $B$-$T$ diagram. It can be seen from Fig. \ref{fig1}\textbf{f} that the $B_{\textrm{c}2}$, traced by the brown diamonds, forms the upper boundary of the half-dome region and exhibits two key features: (1) it exceeds $B_{\textrm{P}}$ at low temperatures; (2) it shows a pronounced upturn as $T\rightarrow 0$ K. Since a low temperature upturn in $B_{\textrm{c}2}$ is a hallmark signature of the FFLO phase~\cite{Fulde01,Larkin01}, these observations indicate that the half-dome region hosts an FFLO phase. This is consistent with previous theoretical studies that an in-plane magnetic field stabilizes a finite momentum pairing in Rashba superconductor~\cite{Gorkov02,Kaur,Feigelman}. Crucially, as the stable solution of the finite momentum pairing in Rashba superconductors has been demonstrated to be the FF phase~\cite{Kaur,Feigelman}, we speculate that the exotic superconducting state in the half-dome region is also the FF phase.

Our speculation about the FF phase in the half-dome region is further supported by the second harmonic resistance measurements at $T<T_{\textrm{c}}$. At $T=3$ K, the finite second harmonic resistance obtained under a $y-$directional magnetic field around $B_y=\pm10$ T (Fig. \ref{fig1}\textbf{h}) reveals a rectification effect~\cite{Itahashi} along the $x$-direction in the half-dome region. Such rectification in the superconducting region is a precursor to the superconducting diode effect~\cite{Wakatsuki2017,Teruo,Itahashi2024}. This observation aligns with recent theoretical works demonstrating that magnetic field induced FF pairing in Rashba superconductors can generate an intrinsic superconducting diode effect~\cite{Daido,Hejun,NoahYuan2022}. Collectively, both the characteristic upturn of $B_{\textrm{c}2}$ and the second harmonic resistance measurements provide compelling evidence that the superconducting state in the emergent half-dome region is the FF phase.

\subsection{FF phase with different NbSe$_2$ thickness}\label{subsec2}


\begin{figure*}[htbp]
	\centering
	\includegraphics[width=1\textwidth,clip]{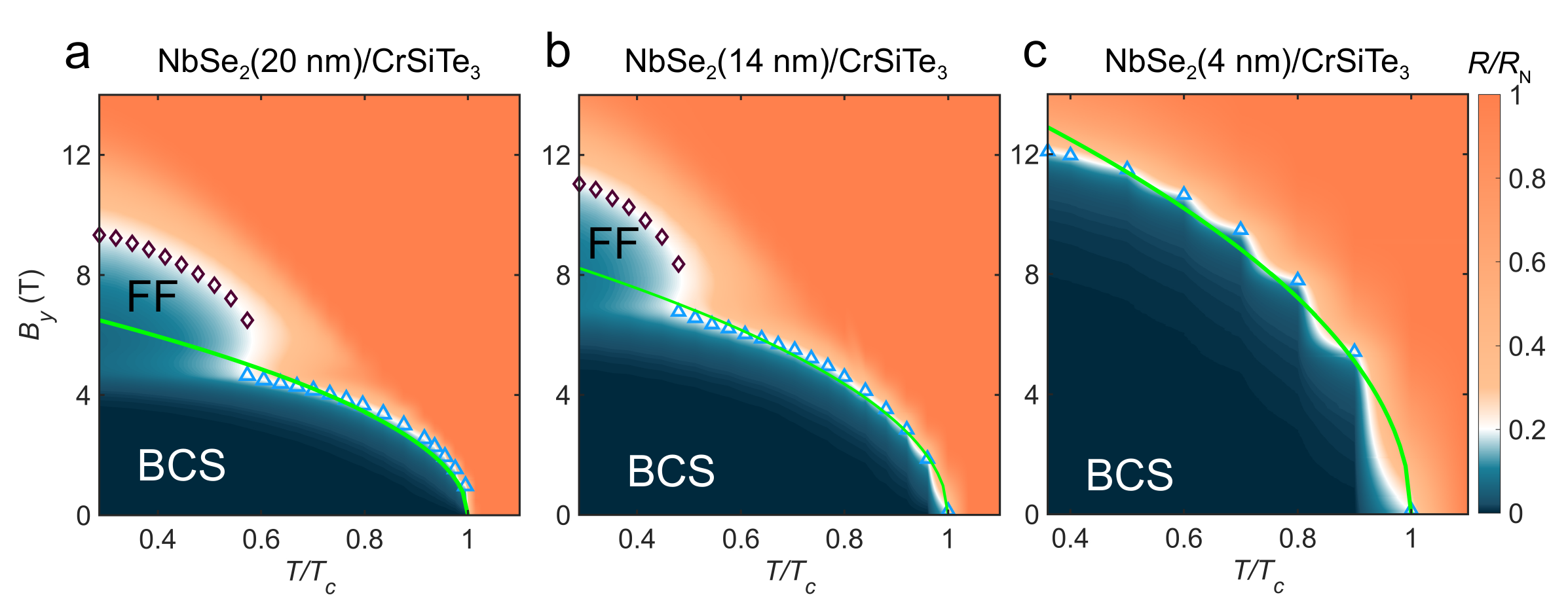}
	\caption{\textbf{FF phase of NbSe$_2$/CrSiTe$_3$ with different thickness of NbSe$_2$.} \textbf{a} - \textbf{c}, $B$-$T$ phase diagrams of the normalized magnetoresistance $R/R_\textrm{N}$ for the NbSe$_2$/CrSiTe$_3$ heterostructures with different thickness of NbSe$_2$ as 20, 14 and 4 nm, respectively. The blue triangles and brown diamonds labeled the $B_{\textrm{c}2}$ in different regions. The green solid lines represent the fitting curves by BCS theory of zero momentum pairing.}\label{fig2}
\end{figure*}


In the superconducting region, the emergent FF phase under an in-plane magnetic field depends critically on the material's specific form of SOC. At the interface of NbSe$_2$/CrSiTe$_3$ heterostructure, both the Ising and Rashba type SOC coexists. In NbSe$_2$, the Ising SOC is highly thickness-dependent, persisting in few-layer flakes while vanishing in the bulk~\cite{XiXiaoxiang2016,WanPuhua2023}, so tuning the thickness of NbSe$_2$ in the heterostructure allows a systematic investigation of the emergent FF phase as the SOC is varied. In addition to the 9 nm NbSe$_2$ sample (Fig. \ref{fig1}\textbf{f}), we fabricated heterostructures with NbSe$_2$ layers of 20 nm (bulk-like), 14 nm and 4 nm (few-layer), coupled to CrSiTe$_3$ with a certain thickness of about 40 nm. By measuring $R/R_\textrm{N}$ as a function of in-plane magnetic field ($B_y$) and temperature ($T$), we obtained the superconducting phase diagrams presented in Fig. \ref{fig2}.


The phase diagram of the NbSe$_2$(20 nm)/CrSiTe$_3$ heterostructure shown in Fig. \ref{fig2}\textbf{a} features two distinct regimes. Above 
$T/T_c\approx0.6$, the upper critical field $B_{\textrm{c}2}$ (blue triangles) can be fitted by the BCS theory for zero momentum pairing. Below this temperature, however, $B_{\textrm{c}2}$ (brown diamonds) exhibits a significant enhancement beyond the BCS fit, forming a half-dome-shaped region identified as the FF phase. As the NbSe$_2$ thickness is progressively reduced to 14 nm, the dimensional crossover enhances the Ising SOC, which consequently narrows the half-dome region toward lower temperatures. With further thinning to 4 nm (Fig. \ref{fig2}\textbf{c}), the intrinsic Ising SOC becomes dominant, leading to the compelete suppression of the half-dome region. The suppression of the half-dome region with decreasing NbSe$_2$ thickness directly demonstrates that the stabilization of the exotic half-dome superconducting state requires Rashba type SOC, rather than Ising SOC. This is consistent with the prevailing understanding that the FF phase is realizable in Rashba superconductors under an in-plane magnetic field~\cite{Gorkov02,Kaur,Feigelman} but is absent in Ising SOC dominated superconductors~\cite{Jianting,de2018Ising,Fai01,Fai02,XingYing2017}.


Our measurements indicate that the NbSe$_2$ thickness serves as a tuning knob for the FF phase, governing its presence in the $B$-$T$ diagram by controlling the SOC at the NbSe$_2$/CrSiTe$_3$ interface. Given that ultrathin NbSe$_2$ (approximately 4 nm or less) hosts a dominant Ising SOC that suppresses the half-dome region, our investigation of the emergent FF phase primarily focuses on heterostructures with thicker NbSe$_2$ layers (exceeding 10 nm).

\begin{figure*}[htbp]
    \centering
    \includegraphics[width=0.9\textwidth,clip]{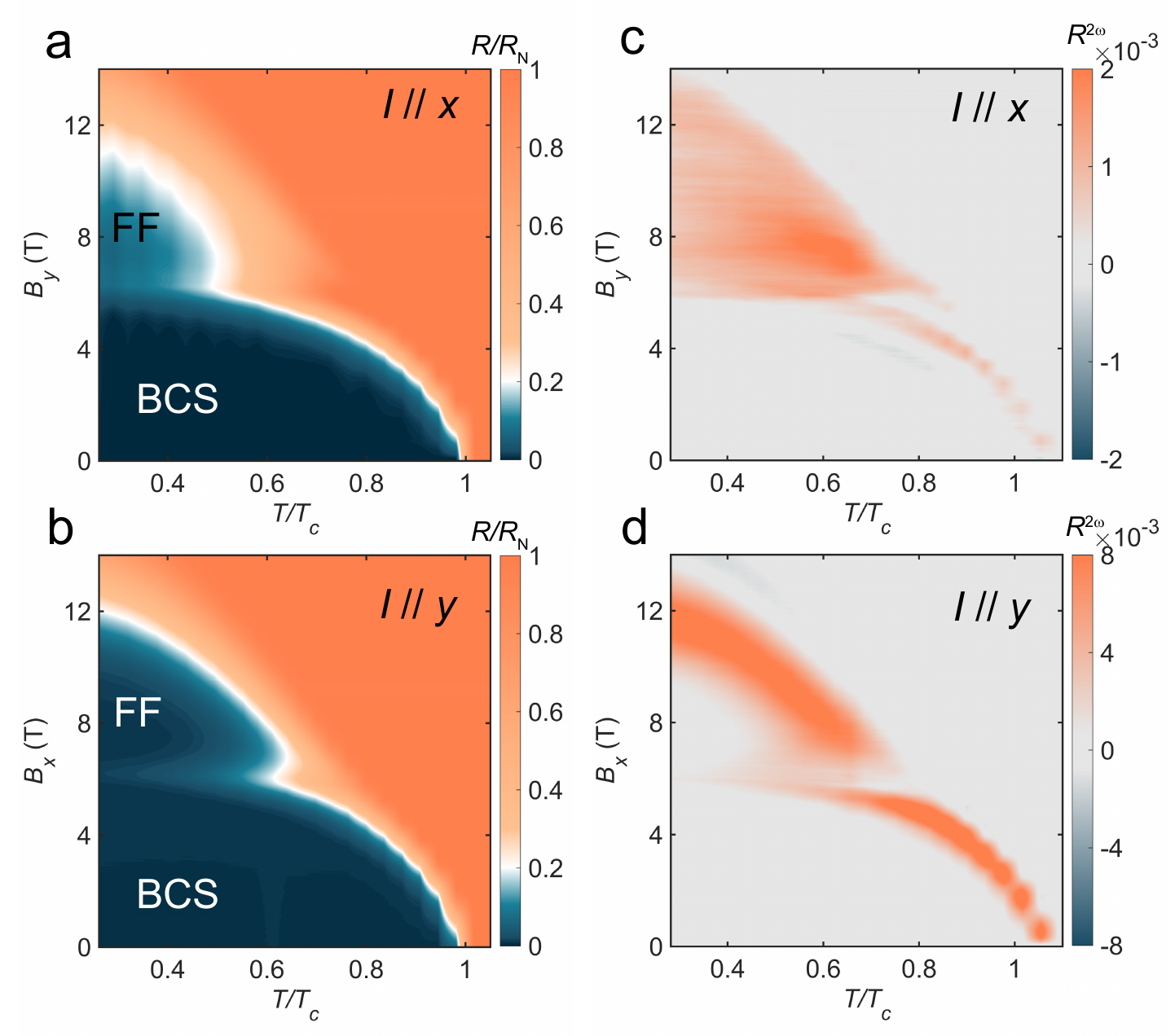}
    \caption{ \textbf{Anisotropic FF phase in NbSe$_2$/CrSiTe$_3$.} \textbf{a}, \textbf{b} $B$-$T$ phase diagrams of $R/R_\textrm{N}$ when the current is applied along the \textit{x}-direction and \textit{y}-direction, respectively. \textbf{c}, \textbf{d}, The corresponding $B$-$T$ phase diagrams of $R^{2\omega}$. During the measurement, the magnetic field is applied in-plane and perpendicular to the current. }\label{fig3}
\end{figure*}

\subsection{The emergent in-plane anisotropy in the FF phase}\label{subsec3}

To systematically investigate the in-plane magnetic field induced FF phase in the NbSe$_2$/CrSiTe$_3$ heterostructure, we further performed transport measurements with the magnetic field applied along different in-plane directions. Surprisingly, our measurements reveal an in-plane anisotropy in the FF phase, which stands in sharp contrast to the presumed isotropy of the underlying Rashba SOC. The in-plane anisotropy is first demonstrated in the $B$-$T$ diagrams in fig. \ref{fig3}\textbf{a} and \ref{fig3}\textbf{b}, where the low temperature $B_{\textrm{c}2}$ differs for magnetic fields applied along the $x$- and $y$-directions. The in-plane anisotropy is further corroborated by the corresponding maps of $R^{2\omega}$ in Fig. \ref{fig3}\textbf{c} and \ref{fig3}\textbf{d}: the finite second hamonic resistance characterizes the FF phase in the half-dome region and exhibits a clear directional dependence. Additionally, our angular-dependent MR measurements with the magnetic field rotating in the \textit{xy}-plane unambiguously show a two-fold anisotropy of the in-plane $B_{\textrm{c}2}$ at low temperatures (see Supplementary Fig. 10). Crucially, the in-plane anisotropy observed in our experiments reflects an intrinsic directional dependence of $B_{\textrm{c}2}$ in the FF phase, which is distinct from the emergent anisotropy arising from the spatial modulation of the pairing gap in the LO phase~\cite{Imajo2022}. Taken together, these findings highlight an intrinsic anisotroy of the FF phase in the emergent half-dome region.

\begin{figure*}[htbp]
    \centering
    \includegraphics[width=0.9\textwidth,clip]{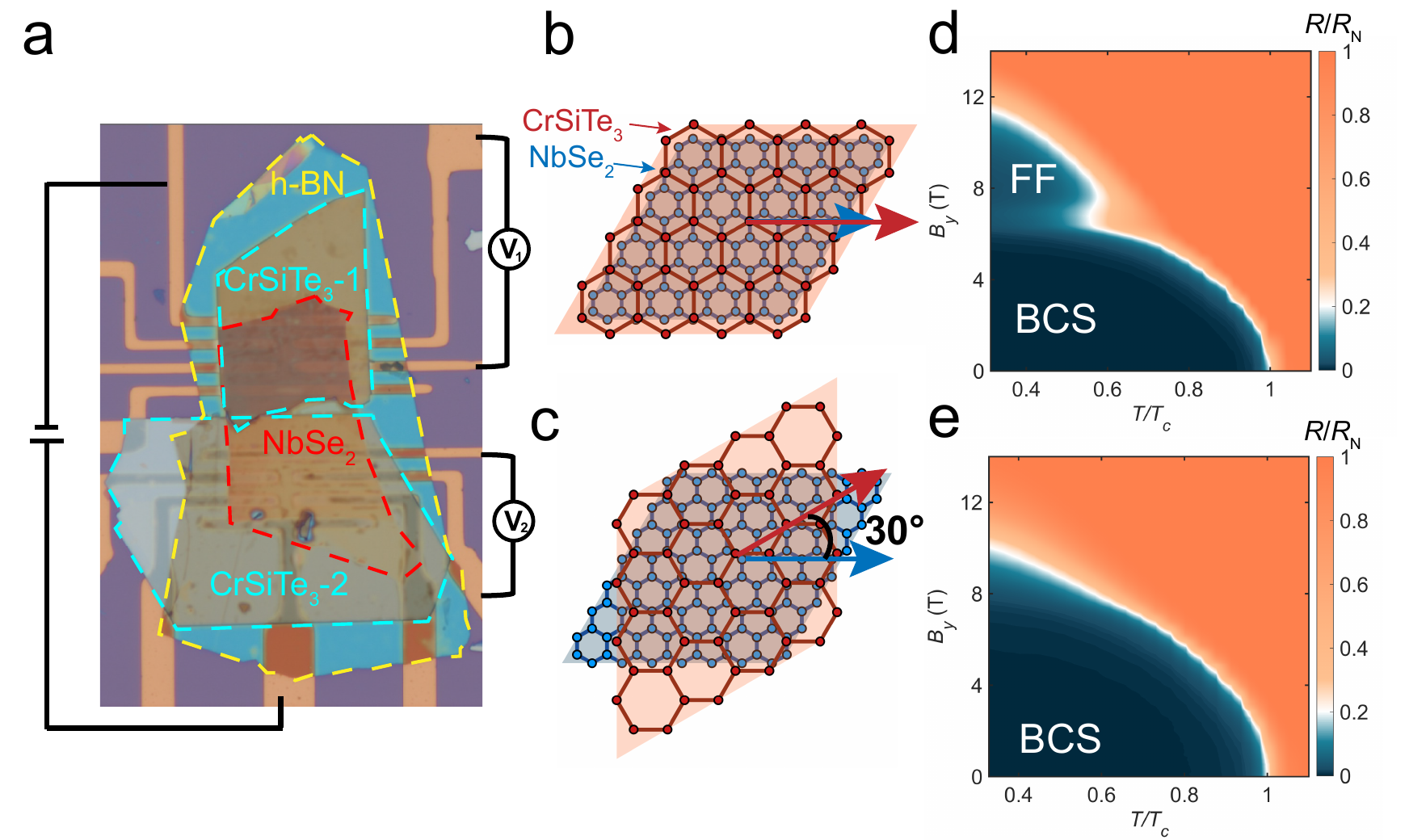}
    \caption{{Twist angle engineering of the FF phase.} \textbf{a} The optical image of the NbSe$_2$/CrSiTe$_3$ heterostructure composed of one NbSe$_2$ layer and two CrSiTe$_3$ layers. The regions outlined by the red, sapphire and yellow dashed boxes correspond to NbSe$_2$, CrSiTe$_3$ and the capping \textit{h}-BN, respectively. \textbf{b}, \textbf{c} The schematic diagrams of the twist angles between NbSe$_2$ and CrSiTe$_3$, which are approximate \textbf{b} 0$^\circ$ and \textbf{c} 30$^\circ$. \textbf{d}, \textbf{e} The corresponding $B$-$T$ phase diagrams of $R/R_\textrm{N}$ for both twist angle configurations.}\label{fig4}
\end{figure*}

\section{Discussion}\label{sec3}

The above results indicate that the superconducting properties of the NbSe$_2$ layers are significantly affected by the presence of CrSiTe$_3$ layer in the heterostructure. Since both NbSe$_2$ and CrSiTe$_3$ have hexagonal lattice structures, their interfacial atomic alignment can be tuned through twist angle engineering. This approach allows us to further investigate how the interfacial atomic registry affects the formation of the FF phase by adjusting the twist angle from 0{\degree} (the most strongly coupling) to 30{\degree} (the most weakly coupling). Figure \ref{fig4}\textbf{a} shows a NbSe$_2$/CrSiTe$_3$ heterostructure consist of one NbSe$_2$ flake in contact with two CrSiTe$_3$ flakes at twist angles of 0{\degree} (lattices aligned, region 1) and 30{\degree} (lattices mismatched, region 2), respectively. Schematic illustrations of the two alignment geometries are presented in Figs. \ref{fig4}\textbf{b} and \ref{fig4}\textbf{c}. The corresponding color maps of the in-plane MR are shown in Figs. \ref{fig4}\textbf{d} and \ref{fig4}\textbf{e}. Notably, signatures of FF phase are observed only in the 0{\degree} twisted flakes, where the atomic alignment is maximized, resulting in the strongest coupling. In contrast, no FF phase signatures are detected in the 30{\degree} twisted flakes, as the atomic alignment is disrupted and the coupling weakens. This suggests that the coupling with CrSiTe$_3$ is essential to stabilize the FF phase in the NbSe$_2$ flake.

\begin{figure*}[htbp]
    \centering
    \includegraphics[width=0.9\textwidth,clip]{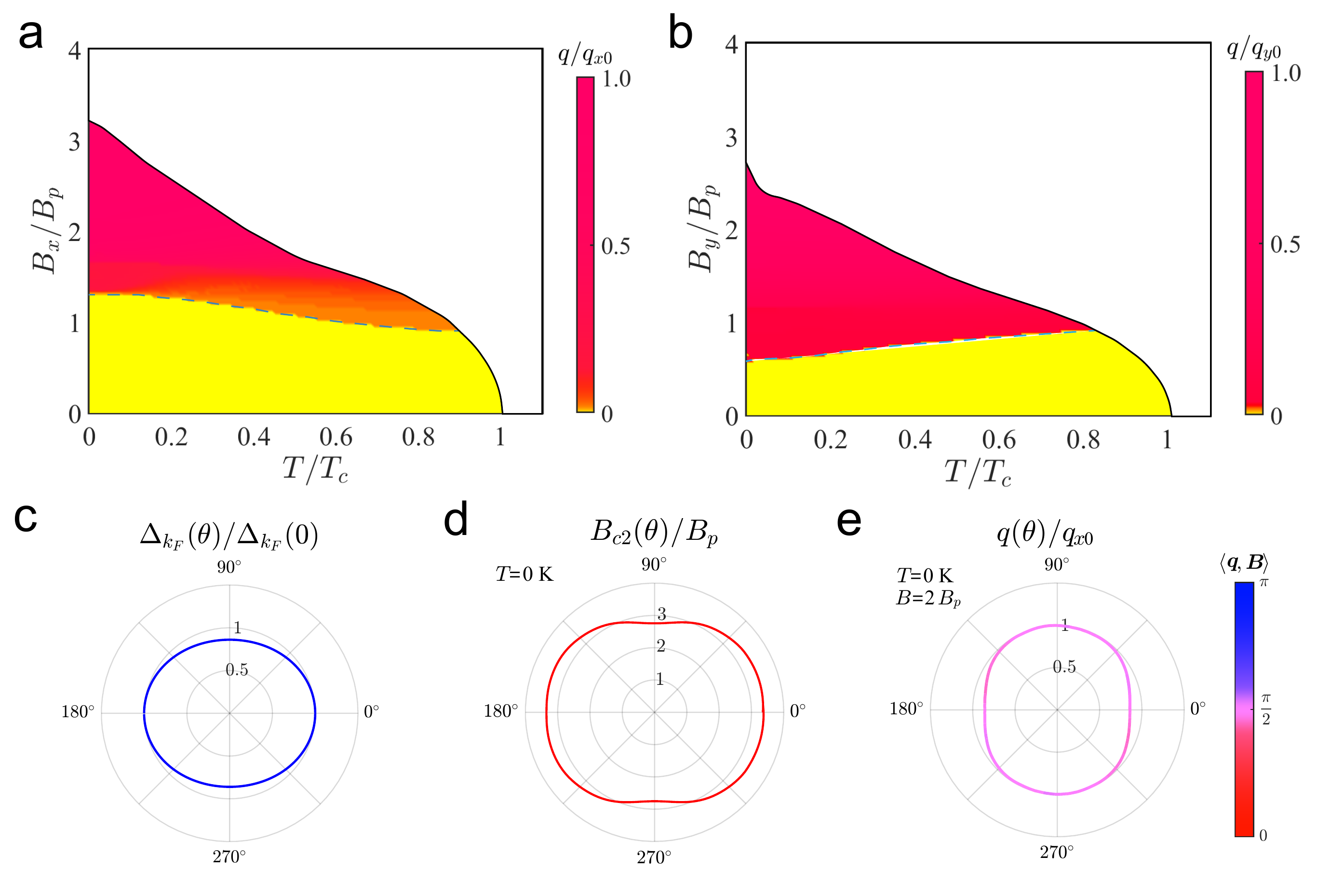}
    \caption{\textbf{Anisotropic FF phase diagram.} \textbf{a} and \textbf{b} The $B$-$T$ phase diagram for the superconducting state with anisotropic FF phase under an in-plane magnetic field along the $x$- and $y$-directions, respectively. At $T=0$K, $x$- and $y$-directional magnetic fields generate finite momentum $\bm{q}=\left(0,q_{x0}\right)$ and $\bm{q}=\left(q_{y0},0\right)$, respectively. \textbf{c} The two-fold asymmetric pairing gap in the $\Gamma$ Fermi pocket. \textbf{d} The polar pllot of the in-plane upper critical field $B_{\textrm{c}2}$. The angle $\theta$ denotes the angle between the in-plane magnetic field and the $x$-axis. \textbf{e} The finite momentum $\bm{q}$ as a function of the in-plane magnetic field $|\bm{B}|=2B_{\textrm{P}}$ with varying in-plane directions. The colorbar $\braket{\bm{q}, \bm{B}}$ denotes the angle between the $\bm{q}$ and the applied in-plane $\bm{B}$. Due to the SOC of C$_{1v}$ symmetry group, $\bm{q}$ is mainly perpendicular to $\bm{B}$. }\label{ff_phasediagram}
\end{figure*}

To understand our experimental observations, we analyze the CrSiTe$_3$ induced coupling effects from the perspective of symmetry breaking. Originally, few layers of NbSe$_2$ in the device exhibits $C_{3v}$ symmetry, with the inversion symmetry broken in both the in-plane and out-of-plane directions, resulting in the coexistence of the Ising and Rashba SOC. In pristine few-layered NbSe$_2$ layers, the in-plane $B_{\textrm{c}2}$ exceeding the Pauli paramagnetic limit indicates that Ising SOC predominates over Rashba SOC~\cite{Fai01,Cuizu}. In NbSe$_2$/CrSiTe$_3$ heterostructures, the NbSe$_2$ layer in contact with CrSiTe$_3$ undergoes additional symmetry breaking. Firstly, the contacted CrSiTe$_3$ further disrupts the out-of-plane inversion symmetry, weakening the Ising SOC while strengthening the Rashba SOC in the NbSe$_2$ layers. As Rashba SOC becomes dominant, Cooper pairs tend to acquire finite momentum to counterbalance the Fermi surface shift under an in-plane magnetic field~\cite{Gorkov02, Sigrist03, Kaur, Feigelman}. Secondly, since CrSiTe$_3$ is a Mott-type ferromagnetic insulator with large magnetic anisotropy~\cite{GangLi}, its in-plane magnetization breaks the three-fold rotational symmetry of the NbSe$_2$ layer, thus inducing in-plane anisotropy in the FF phase.

Guided by symmetry analysis, we performed mean field calculations for the $B$-$T$ diagram of the NbSe$_2$ layer in contact with CrSiTe$_3$. Supplementary Fig. 17 illustrates the effect of Rashba SOC on the $B$-$T$ diagram of the NbSe$_2$ layer. With increasing Rashba SOC, an in-plane magnetic field applied at low temperatures induces a transition of the superconducting state into the FF phase. For weaker Rashba SOC, the FF phase is suppressed. The results presented in Supplementary Fig. 17 align with our analysis that interfacial coupling to CrSiTe$_3$ strengthens the Rashba SOC in the NbSe$_2$ layer, facilitating the formation of the FF phase. To account for the three-fold rotational symmetry breaking, we assume that the point group of the NbSe$_2$ layer is reduced from $C_{3v}$ to $C_{1v}$, with one in-plane mirror plane retained. This is compatible with the spontaneous nematic pairing~\cite{LiangFu,LiangFu2} (also see supplementary note 4). For the NbSe$_{2}$ layer with reduced symmetry to $C_{1v}$, anisotropy arises in both the pairing gap and SOC (see supplementary Note 4), leading to distinct responses to in-plane magnetic fields applied along the $x$ and $y$ directions. As shown in Fig. \ref{ff_phasediagram}, our simulations reveal significant in-plane directional anisotropy: the upper critical field ($B_{\textrm{c}2}$), the threshold field for FF phase onset, and the finite momentum of the pairing state all differ for $x$ and $y$ directional magnetic fields. This in-plane anisotropy is further evidenced by the emergent two-fold symmetry in both the in-plane $B_{\textrm{c}2}$ and the finite momentum of the pairing state (Figs. \ref{ff_phasediagram}\textbf{d} and \ref{ff_phasediagram}\textbf{e}). These findings are fully consistent with our experimental observations of anisotropic FF phase behavior in NbSe$_2$/CrSiTe$_3$ heterostructures.  

\section{Conclusion}\label{sec4}

In summary, we have observed an anisotropic FF phase in NbSe$_2$/CrSiTe$_3$ heterostructures under in-plane magnetic fields at low temperatures. By integrating symmetry analysis and mean field calculations, we establish that the emergent FF phase and its in-plane anisotropy originate from proximity-induced symmetry breaking by CrSiTe$_3$. These findings advance our understanding of the pairing states in NbSe$_2$ under in-plane magnetic fields while demonstrating how heterostructure engineering can tailor the pairing states. The heterostructure stacking technique demonstrated in our work provides a versatile strategy to engineer superconducting symmetries in atomically thin materials. Beyond fundamental insights into unconventional superconductivity, this approach also opens avenues for designing superocnducting diode devices~\cite{Teruo, Faxian, Nicola} and realizing topological superconductivity~\cite{Liljeroth, Wenyu01, Xinzhi}, bridging the gap between symmetry driven quantum phenomena and functional superconducting technologies.

\backmatter

\section*{Methods}\label{sec5}

\subsection*{Device fabrications} \label{subsec4}

It started from mechanically exfoliating van der Waals CrSiTe$_3$ on Si/SiO$_2$ substrate. Then, NbSe$_2$ was mechanically exfoliated on polydimethylsiloxane (PDMS) and dry-transferred on the surface of CrSiTe$_3$ flake. Finally, Hexagonal boron nitride ($h$-BN) is used for encapsulation and protecting the device from degeneration. The circle pattern was written by Laser Direct-write lithography system, after reactive ion etching, 5 nm Ti/30 nm Au was electron beam deposited and directly contacted with NbSe$_2$ layer.

To minimize the contact resistance and to avoid interfacial contamination, both NbSe$_2$ and CrSiTe$_3$ were exfoliated and transferred onto a clean Si/SiO$_2$ substrate and encapsulated by a protective layer $h$-BN in a glove box filled with Ar atmosphere. Ti/Au contact electrodes were deposited into the grooves on the surface of $h$-BN etched by reactive ion etching technique (See Supplementary Note 1). For comparison, the NbSe$_2$/CrSiTe$_3$ heterostructure and the bare NbSe$_2$ flake were designed on the same sample. Moreover, a few-layer graphene was inserted into the heterostructure to reduce the interfacial Rashba effect. 

\subsection*{Transport measurements} \label{subsec5}

The transport measurement was carried out in PPMS. The four-terminal DC and AC signal was measured by set of Keithley 2400 and 2182a and set of Keithley 6221 and OE1022 lock-in amplifier, respectively. The first and second harmonic resistance were defined as $R^{\omega}=V^{\omega}/I_{0}$ and $R^{2\omega}=V^{2\omega}/I_{0}$, where $I_{0}$ is the amplitude of the AC current applied and $V^{\omega}$ and $V^{2\omega}$ are the amplitude of first and second harmonic voltage. The current frequency was set to be 113 Hz to lower the noise and the phase of second harmonic signal was set as $\pi/2$.

\subsection*{Mean field calculations}\label{subsec6}
After establishing an effective two-band tight binding model for a NbSe$_2$ layer coupled with CrSiTe$_3$ (See Supplementary Note 3), the general Hamiltonian that involves a BCS type pairing attractive interaction can be written as
\begin{align}\nonumber
	\mathcal{H}&=\sum_{\bm{k},i,s,s'}c^{\dagger}_{\bm{k},s}[h_{s,s'}(\bm{k})-\mu+\frac{1}{2}g\mu_{\textrm{B}}B_i\sigma_{i,ss'}]c_{\bm{k},s'}\nonumber\\
	&-\dfrac{U}{\Omega}\sum_{\bm{k},\bm{k}',\bm q}c^{\dagger}_{\bm{k}+\frac{\bm q}{2},\uparrow}c^{\dagger}_{-\bm{k}+\frac{\bm q}{2},\downarrow}c_{-\bm{k}'+\frac{\bm q}{2},\downarrow}c_{\bm{k}'+\frac{\bm q}{2},\uparrow},
\end{align}
where $\Omega$ denotes the volume of the sample, $c^{\dagger}_{\bm k,s}$ is the creation operator and $h_{s,s'}\left(\bm{k}\right)$ denotes the Hamiltonian matrix for the NbSe$_{2}$ normal state. Here $s=\uparrow/\downarrow$ represents the spin index, and $i=x,y,z$ denotes the spatial components of magnetic field. We consider a constant effective attraction $U$ and adopt a generalized pairing configuration including finite pairing momentum $\bm q$. Within the mean field approximation, the partition function for the superconducting pairing state is derived via the path integral formalism~\cite{altland2010condensed} 
\begin{align}\nonumber
	\mathcal{Z}\approx&\int \mathcal{D}\left[\psi^{\dagger}_{\bm{k},\bm q},\psi_{\bm{k},\bm q}\right]\exp\Bigl\{-\frac{\beta\Omega|\Delta|^2}{U}\nonumber\\
	&+\frac{1}{2}\sum_{\bm{k},n}\Psi^{\dagger}_{\bm{k},\bm q,n}\left[i\omega_n-H_{\bdg}(\bm{k},\bm q)\right]\Psi_{\bm{k},\bm q,n}\Bigr\}\nonumber\\
	=&\exp\Bigl\{-\frac{\beta\Omega|\Delta|^2}{U}+\frac{1}{2}\sum_{\bm{k},n}\tr\log\left[-\beta G^{-1}(\bm{k},\bm q,i\omega_n)\right]\Bigr\}.
\end{align}
Here $\Psi^{\dagger}_{\bm{k},\bm q}=\left[c^{\dagger}_{\bm{k}+\frac{\bm q}{2},\uparrow},c^{\dagger}_{\bm{k}+\frac{\bm q}{2},\downarrow},c_{-\bm{k}+\frac{\bm q}{2},\uparrow},c_{-\bm{k}+\frac{\bm q}{2},\downarrow}\right]^{\textrm{T}}$ is the Nambu spinor, $G^{-1}(i\omega_n,\bm k,\bm q)=i\omega_n-H_{\bdg}(\bm k,\bm q)$ is the Matsubara Green's function, and $H_{\bdg}(\bm k,\bm q)$ is the Bogliubov de Gennes Hamiltonian
\begin{align}\label{h_BdG}
	H_{\bdg}\left(\bm{k}\right)=\begin{pmatrix}
	\tilde{h}\left(\bm{k}+\frac{\bm q}{2}\right)-\mu & -i\sigma_y\Delta\psi\left(\bm{k}\right) , \\
	\psi^\ast\left(\bm{k}\right)\Delta i\sigma_y & -\tilde{h}^\ast\left(-\bm{k}+\frac{\bm q}{2}\right)+\mu
	\end{pmatrix}.
\end{align}
with $\tilde{h}(\bm k)=h(\bm k)+\frac{1}{2}g\mu_B\bm B\cdot\bm \sigma$ incorporating the Zeeman coupling. The in-plane magnetic field considered in our calculations takes the form $\bm{B}=B\left(\cos\theta,\sin\theta,0\right)$. Here $\Delta$ denotes the pairing amplitude and $\psi\left(\bm{k}\right)$ denotes the function form of the pairing order parameter (see Supplementary Note 4 for details).

The free energy density $F\left(\bm{q}\right)=-\frac{1}{\beta\Omega}\ln\mathcal{Z}$ is then given by
\begin{align}\label{free}
	F\left(\bm{q}\right)=\frac{|\Delta|^2}{U}-\frac{1}{2\beta\Omega}\sum_{\bm k,\nu}\ln\left(1+e^{-\beta\xi_{\nu,\bm k,\bm q}}\right)
\end{align}
with $\xi_{\nu,\bm k,\bm q}$ being the eigenvalues of $H_{\bdg}(\bm k,\bm q)$ in Eq. \eqref{h_BdG}. In the case of zero momentum pairing, the center-of-mass momentum is fixed to be $\bm{q}=\bm{0}$. At $B=0$~T, the superconductor is of zero momentum pairing and the critical temperature is $T_c=7.3$~K. The effective attraction $U$ can then be determined by solving $\frac{\partial F\left(\bm{0}\right)}{\partial \Delta}=0$. We assume that the effective attraction $U$ is unchanged when applying a magnetic field. Under a finite in-plane magnetic field ($B\neq0$ T), since Cooper pairs may acquire finite center-of-mass momentum to further reduce the free energy, one needs to vary both $\Delta$ and $\bm{q}$ to find the global minimal free energy density. The $B$-$T$ phase diagram is thus obtained through numerically minimizing $F\left(\bm{q}\right)$ with respect to both $\Delta$ and $\bm{q}$ at fixed $B$ and $T$.





\section*{Acknowledgements}\label{subsec7}

This research was supported in part by the Ministry of Science and Technology (MOST) of China (No. 2022YFA1603903), the National Natural Science Foundation of China (Grants No. 62571327, 92565112, 92565305), the Science and Technology Commission of Shanghai Municipality, the Shanghai Sailing Program (Grant No. 21YF1429200), Natural Science Foundation of Shanghai (Grants Nos. 25ZR1402374, 25ZR1402368, and 25ZR1401252), the Science and Technology Commission of Shanghai Municipality, the Shanghai Leading Talent Program of Eastern Talent Plan, and the Double First-Class Initiative Fund of ShanghaiTech University. W.-Y.H. acknowledges the support from the National Natural Science Foundation of China (No. 12304200), the BHYJRC Program from the Ministry of Education of China (No. SPST-RC-10), the Shanghai Rising-Star Program (24QA2705400), and the start-up funding from ShanghaiTech University. Growth of hexagonal boron nitride crystals was supported by the Elemental Strategy Initiative conducted by the MEXT, Japan, Grant Number JPMXP0112101001, JSPS KAKENHI Grant Number JP20H00354 and A3 Foresight by JSPS. The authors also thank the nano-fabrication supporting from the Soft Nano Fabrication Center at ShanghaiTech University.

\bibliography{reference}

\end{document}